\documentclass[11pt]{book}

\usepackage{amssymb}
\usepackage{amsmath}
\usepackage{amsthm}
\usepackage{epsfig}
\usepackage{graphicx}
\usepackage[colorlinks=true]{hyperref}

\usepackage[usenames]{color}
\definecolor{rred}{rgb}{1,0,0}
\definecolor{blue}{rgb}{0,0,1}
\definecolor{violet}{rgb}{1,0,1}

\allowdisplaybreaks[1]


\newcommand\vcenterbox[1]{\vcenter{\hbox{#1}}}
\newcommand{\sect}[1]{\setcounter{equation}{0}\section{#1}}

\renewcommand\bibname{References}

\def\oh{{1\over 2}}

\def\Md{M^{\dagger}}

\def\Gl{\lambda}
\def\tG{\tilde{\Gamma}}
\def\tH{\tilde{H}}
\def\E#1{{\rm e}^{\textstyle #1}}
\long\def\rem#1{{\sl [#1]}}

\def\be{\begin{equation}}
\def\ee{\end{equation}}
\def\bbra{\langle}\def\kket{\rangle}
\def\d{{\rm d}}
\newcommand\tr{\mathop{{\rm tr}\over N}\nolimits}
\newcommand\Tr{\mathop{\rm tr}\nolimits}
\newtheorem*{Theorem}{Theorem} 
\newtheorem*{Problem}{Problem} 
\begin{document}

\setlength{\baselineskip}{5.0mm}



\chapter[Knot theory -- Zinn-Justin and Zuber]{Knot theory and matrix integrals}
\thispagestyle{empty}

\ \\

\noindent
{{\sc Paul Zinn-Justin} and {\sc Jean-Bernard Zuber}
\\~\\ UPMC Univ Paris 6, CNRS UMR 7589, LPTHE \\ 75252 Paris Cedex}

\begin{center}
{\bf Abstract}
\end{center}
The large size limit of matrix integrals with quartic potential 
may be used to count alternating links and tangles. 
The removal of redundancies amounts to renormalizations of the potential. 
This extends into two directions: higher genus and the counting of ``virtual"
links and tangles; and the counting of ``coloured'' alternating links and tangles.
We discuss the asymptotic behavior of the number of tangles as the number of crossings
goes to infinity.


\sect{Introduction and basic definitions}\label{basics}
This chapter is devoted to some
enumeration problems in knot theory. For a general review of the subject, 
see \cite{Finch-knots}. Here we are interested in the application of matrix integral techniques.
We start with basic definitions of knot theory.

\subsection{Knots, links and tangles}
We first recall the definitions of the knotted objects under consideration. 
A {\it knot} is a closed loop embedded in 3-dimensional space. 
A {\it link} is made of several  entangled knots. An $n$-{\it tangle} is
a knotted pattern with $2n$ open ends. We shall be interested in particular
in 2-tangles, where it is conventional to attach the four outgoing strands
to the four cardinal points SE, SW, NW, NE.\quad

\includegraphics[width=.8\textwidth]{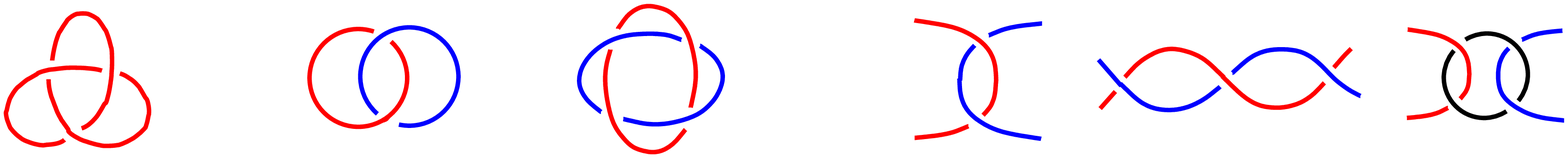}\\
This figure depicts a knot, two links  and three 2-tangles.  

All these objects are regarded as equivalent under 
isotopy {\it i.e.} under 
deformations in which strands do not cross one another, 
and (for tangles) open ends are maintained fixed.
Our problem is to count topologically inequivalent knots, links and tangles.

It is usual to represent knots etc by their {\it planar projection} 
with minimal number of over/under-crossings. There is an important 
\begin{Theorem}[Reidemeister]
Two projections represent the same knot, link or tangle iff 
they may be transformed into one another by a sequence of 
Reidemeister moves:
\raise-7mm\hbox{\includegraphics[width=.75\textwidth]{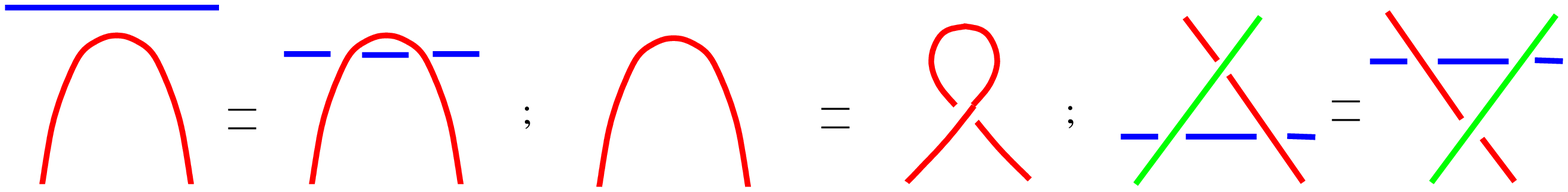}}
\end{Theorem}

\noindent \begin{minipage}[t]{0.8\textwidth}
Also, in the classification or the counting of knots etc, one tries to
avoid redundancies by keeping only
{\it prime} links. 
A link is non prime if cutting tranversely two strands may yield 
 two disconnected non trivial parts. Here is a non prime link:
\end{minipage} \hfill
\begin{minipage}[t]{0.18\textwidth}\vspace{20pt}
\centering \includegraphics[width=0.9\textwidth]{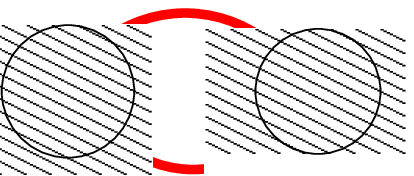}
\end{minipage}


\subsection{Alternating links and tangles}

We shall now restrict ourselves to the subclass of {\it alternating} knots, links and tangles,
in which one meets alternatingly over- and under-crossings, when one follows any strand.

\noindent  \begin{minipage}[t]{0.83\textwidth}
\quad\  For low numbers of crossings, all knots, links or tangles may be drawn in an alternating 
pattern, but 
for $n\ge 8$ (resp. 6) crossings, there are knots (links)
which cannot be drawn in an alternating form.
Here is an example of a 8-crossing non-alternating knot:
\end{minipage} \hfill
\begin{minipage}[t]{0.15\textwidth}\vspace{0pt}
\centering \includegraphics[width=0.8\textwidth]{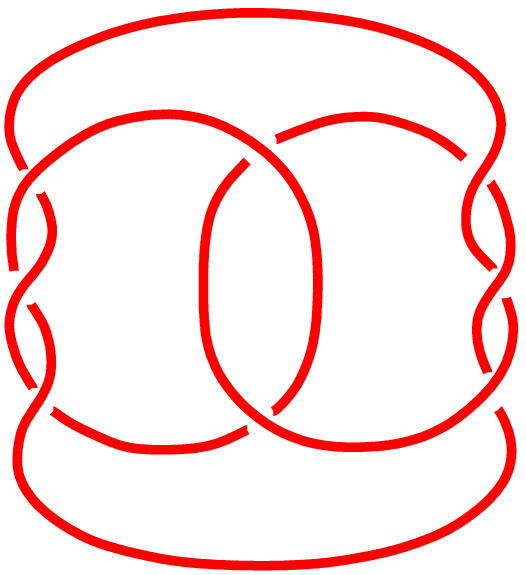} 
\end{minipage}

One may show that 
asymptotically, the alternating links and knots are subdominant. Still the tabulation and counting
of this subclass is an important task, as  a preliminary step in the general
classification program. 

A major result conjectured by Tait (1898) and proved 
in \cite{MT-flype,MT-altlinks}~is~the 
\begin{Theorem}[Menasco, Thistlethwaite]
Two alternating reduced knots or links represent  the same object
iff they are related by a sequence of {\it ``flypes''},

\noindent \begin{minipage}[t]{0.45\textwidth}
where a  flype is a combination of Reidemeister moves respecting the alternating character 
of tangles:
\end{minipage} \hfill
\begin{minipage}[t]{0.55\textwidth}\vspace{0pt}
\centering \includegraphics[width=0.9\textwidth]{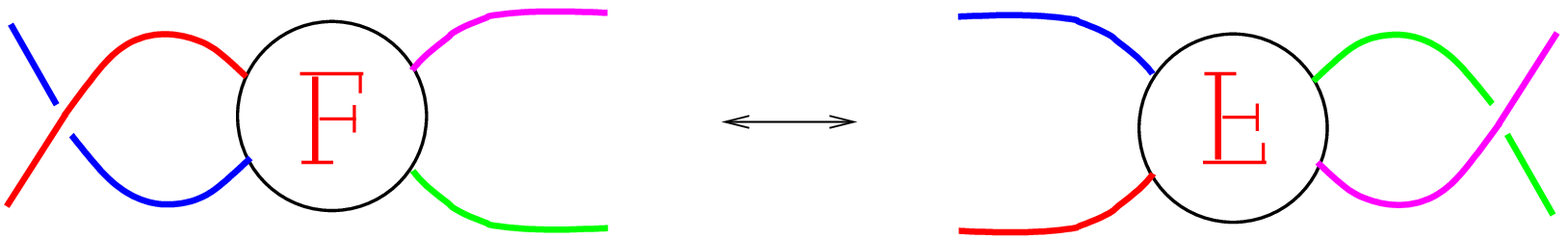}
\end{minipage}
\end{Theorem}

We shall thus restrict ourselves  to the (manageable)\nobreak
\begin{Problem}
Count alternating prime links and tangles.
\end{Problem}
This problem was given a first substantial answer by Sundberg
      and Thistlethwaite in \cite{ST-tangles}. We will discuss in the rest of
       this text how the matrix integral approach has allowed to make
       significant progress building on their work.


\sect{Matrix integrals and\,alternating links\,and\,tangles}\label{matrint}

\subsection{The basic integral}\label{sect131}
Consider the  integral over complex ({\it non Hermitean}) $N\times N$ matrices

\be
 Z_C= \int dM\, \E{N[-{t} \, \tr M \Md +{g\over 2} \tr (M \Md)^2]}
\label{intc}
\ee
with $dM=\prod_{i,j}d\Re e M_{ij}\,d\Im m M_{ij}$. It was proposed in the
context of knot enumeration in \cite{artic08}.

\noindent \begin{minipage}[t]{0.75\textwidth}
\quad \ According to the discussion of Chapter \rem{?}, its diagrammatic expansion involves oriented 
double-line propagators \includegraphics[width=.20\textwidth]{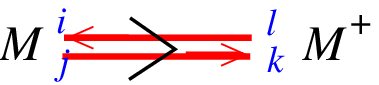},  while its vertices
may be drawn in a one-to-one correspondence with the previous link crossings, 
with, say,  over-crossing associated with outgoing arrows.
\end{minipage} \hfill
\vspace{.2cm}
\begin{minipage}[t]{0.20\textwidth}\vspace{0pt}
\centering \includegraphics[width=0.9\textwidth]{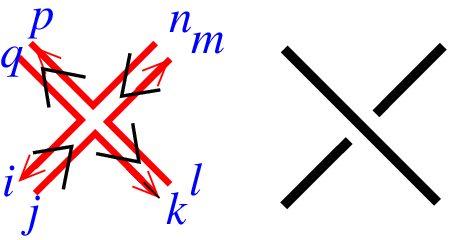}
\end{minipage}

As usual, the perturbative (small $g$) expansion of the integral (\ref{intc}) or of the associated correlation
functions involves only planar diagrams in the large $N$ limit. 
Moreover the conservation of arrows implies that the diagrams are alternating:\\
\hskip2cm \centerline{\includegraphics[width=.35\textwidth]{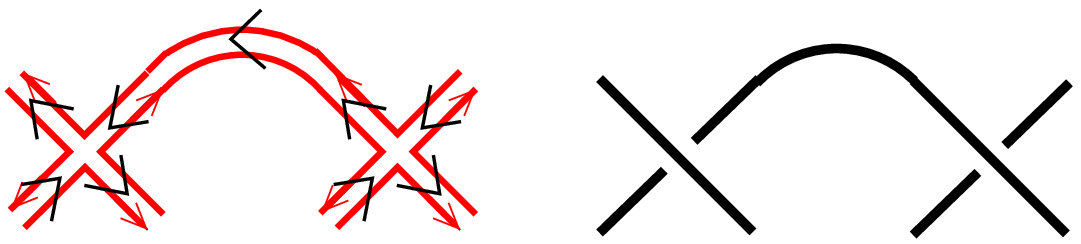}}

It follows from the discussion of Chapter XX that in the large $N$ limit
{$\lim_{N\to\infty} {1\over N^2} \log Z_C = 
\sum_{{\rm planar\ connected\ alternating} 
\atop {\rm diagrams\ } D\ {\rm  with\ } n\ { \rm  vertices}} {g^n\over |{\rm Aut} D|}$}, 
where $ |{\rm Aut} D|$ is the order of the automorphism group of $D$. 
%
But going from complex matrices to hermitian matrices
 doesn't affect that ``planar limit'',  
 up to a global factor 2. We thus conclude that, 
provided we remove redundancies including flypes, the counting of Feynman diagrams
of the following integral over 
$N\times N$ Hermitean matrices $M$, for $N\to\infty$, 
\be
 Z=\int dM\, \E{N[-{{t}\over 2} \tr M^2 +{g\over 4} \tr M^4]} 
\label{inth}
\ee
with $dM=\prod_i dM_{ii} \prod_{i<j}d\Re e M_{ij}\,d\Im m M_{ij}$, 
yields the counting of 
alternating links and tangles. 


\subsection{Computing the integral} 
The large $N$ limit of the integral (\ref{inth})  may be computed 
by the saddle point method, by means of orthogonal polynomials or 
of the loop equations, as reviewed elsewhere in this book.

In that $N\to \infty$ limit, the eigenvalues $\Gl$ form a 
 continuous distribution with 
density  $u(\Gl)$ of support $[-2a,2a]$, forming a deformed semi-circle law \cite{BIPZ}\\
\be
{u(\Gl)={1\over 2\pi}(1-2{g\over t^2}a^2-{g\over t^2}
\Gl^2)\sqrt{4a^2-\Gl^2}}
\ee
with $a^2$ related to $g$ and $t$ by 
\be
{3{g\over t^2} a^4 -a^2 +1=0 }
\ee
and one finds that the large $N$ limit of the ``free energy" $F$ is
\begin{align*}
F(g,t)& :=\lim_{N\to\infty} {1\over N^2}\log {Z(g,t)\over Z(t,0)}=
\oh \log a^2-{1\over 24}(a^2-1)(9-a^2)\\ 
F(g,t) &=\sum F_p \left({g\over t^2}\right)^p= \sum_{p=1} \Big({3g\over t^2}\Big)^p {(2p-1)!\over p!(p+2)!} \,.
\end{align*}
We recall that this formal power series of $F$, the  ``perturbative expansion of $F$'', 
 is a generating function for the number
of {\it connected} planar diagrams, (as usual, weighted by their inverse symmetry factor)\\
 \includegraphics[width=\textwidth]{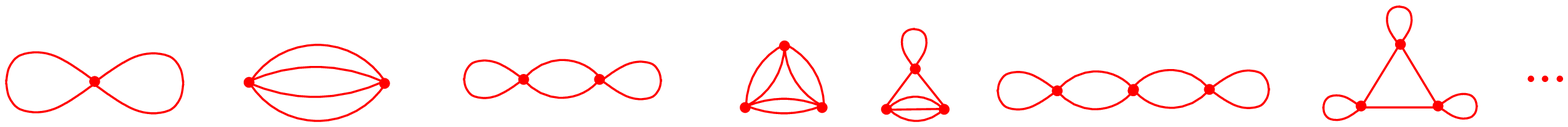}
For future reference, we note that the asymptotic behavior of $F_p$ as
 $p\to\infty$ is 
 \be  F_p\sim 
{\rm const} (12)^p p^{-7/2} \,.
\label{Fasymp1}
\ee

Also all the 2$p$-point functions $\frac{1}{N}\bbra \tr  M^{2p}\kket=\int \rho(\Gl) \Gl^{2p}$ may be computed.
We only give here two expressions that we need below, the 2-point function
\be
 \Delta={1\over 3 t}a^2(4-a^2) = \vcenterbox{
\includegraphics[width=.13\textwidth]{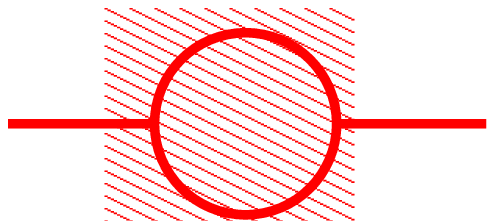}}
\label{2ptfn}
\ee
%
and the  connected 4-point function $\Gamma$ 
\be
 \Gamma(g,t)={1\over 9 t^2}a^4(1-a^2)(2a^2-5) =\vcenterbox{
\includegraphics[width=.09\textwidth]{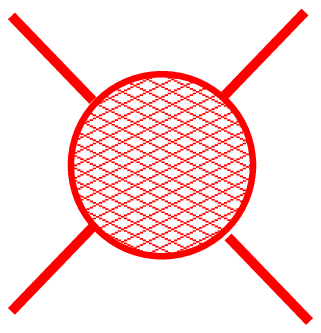}} ,
\label{2ptfnfig}
\ee
whose diagrams, after removal of redundancies, will count 2-tangles.  The $p$-th term $\gamma_p$ in the 
$g$ expansion of $\Gamma$ behaves as $12^p p^{-5/2}$. 

\subsection{Removal of redundancies}\label{sect133}

The removal of redundancies for the counting links and  tangles
will be done in two steps. 
First ``nugatory'' that are in fact irrelevant diagrams representing patterns that may be unknotted, 
 $\vcenterbox{ \includegraphics[width=.2\textwidth]{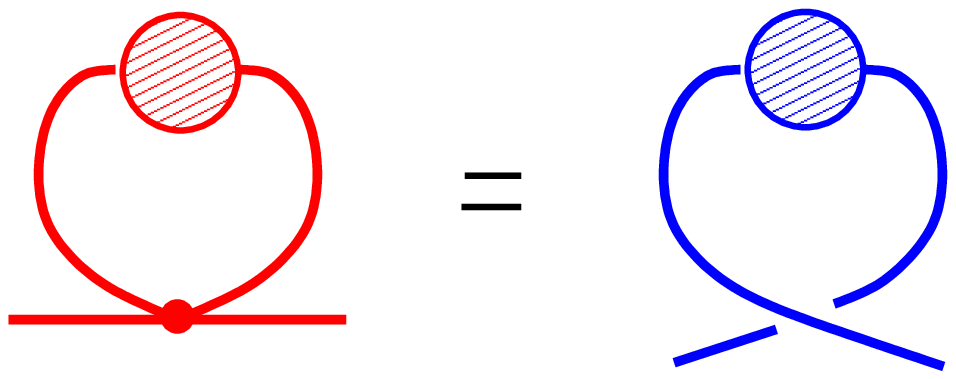}}$\,, 
and ``non-prime'' diagrams
 $\vcenterbox{\includegraphics[width=.1\textwidth]{fig3.eps}}$, may be both 
removed by adjusting 
$t=t(g)$ 
in such a way 
that $\Delta=\vcenterbox{\includegraphics[width=.1\textwidth]{twopoint.eps}}
=1$. In the language of quantum field theory, this is a ``wave function renormalisation''.

We then find $F(g)=F(g,t(g))$ 
\be
F(g) = \frac{g^2}{4} +\frac{g^3}{3} +\frac{3g^4}{4}+ \frac{11 g^5}{5}+ \frac{91g^6}{12} +\cdots 
\ee
and $\Gamma(g)=\Gamma(g,t(g))={(5-2 a^2)(a^2-1)\over (4-a^2)^2} =2 \frac{\d F}{\d g}$.

In that way, one gets the correct counting of links  up to 6 crossings and of 2-tangles up to 3 crossings.
This is apparent  on the following table where we have listed in (a) the first links with their 
traditional nomenclature; (b) the corresponding Feynman diagrams with their symmetry weight;
(c) the first 2-tangles in Feynman diagram notation. It appears that the diagrams  of
each of the last two pairs in (c) are flype equivalent. 
 
\includegraphics[width=\textwidth]{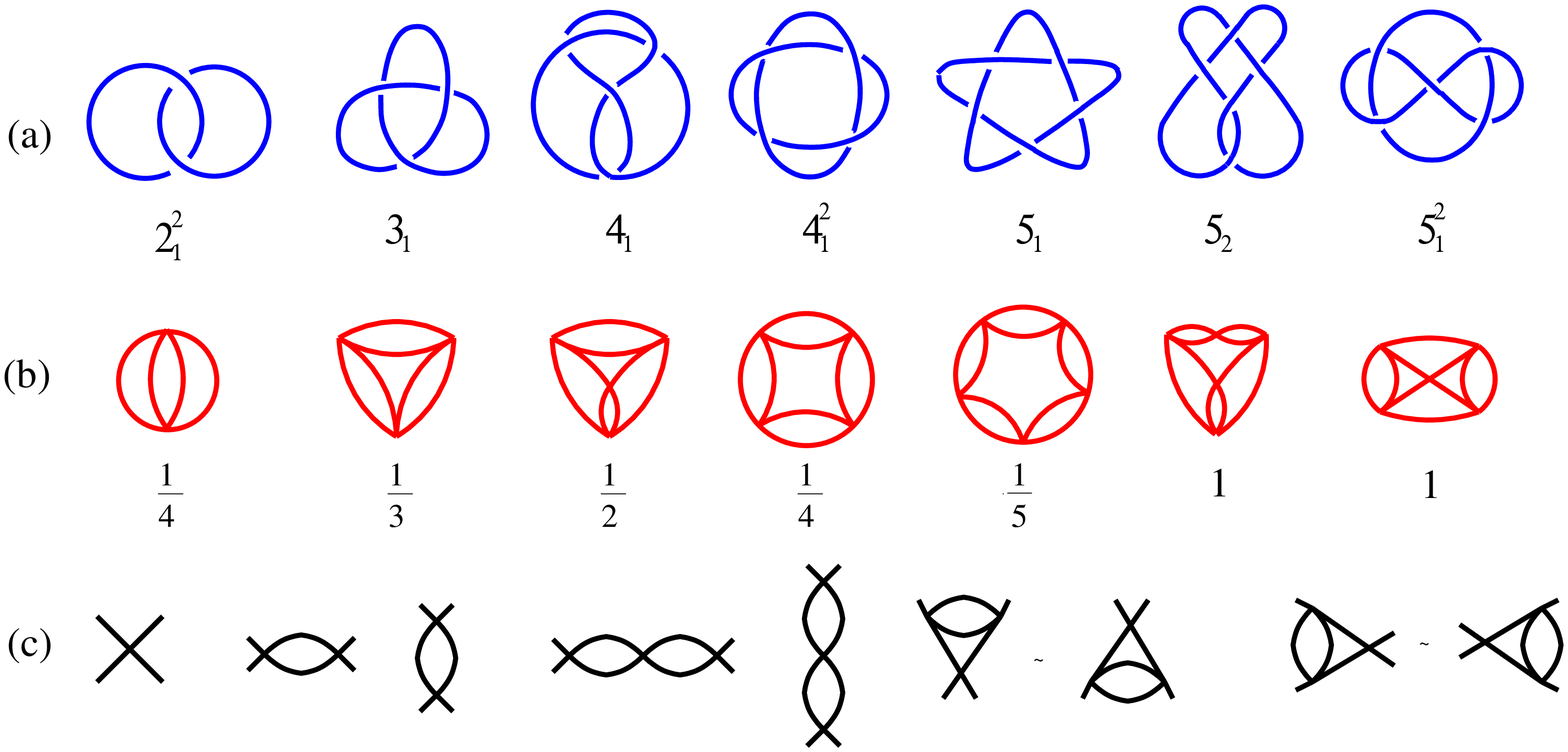} 

\bigskip
\hskip-5mm\begin{minipage}[t]{0.45\textwidth}
\noindent For links the first flype equivalence occurs at order 6: 
\end{minipage} \hfill
\vspace{-.4cm}
\begin{minipage}[t]{0.35\textwidth}\vspace{0pt}
\centering\includegraphics[width=0.85\textwidth]{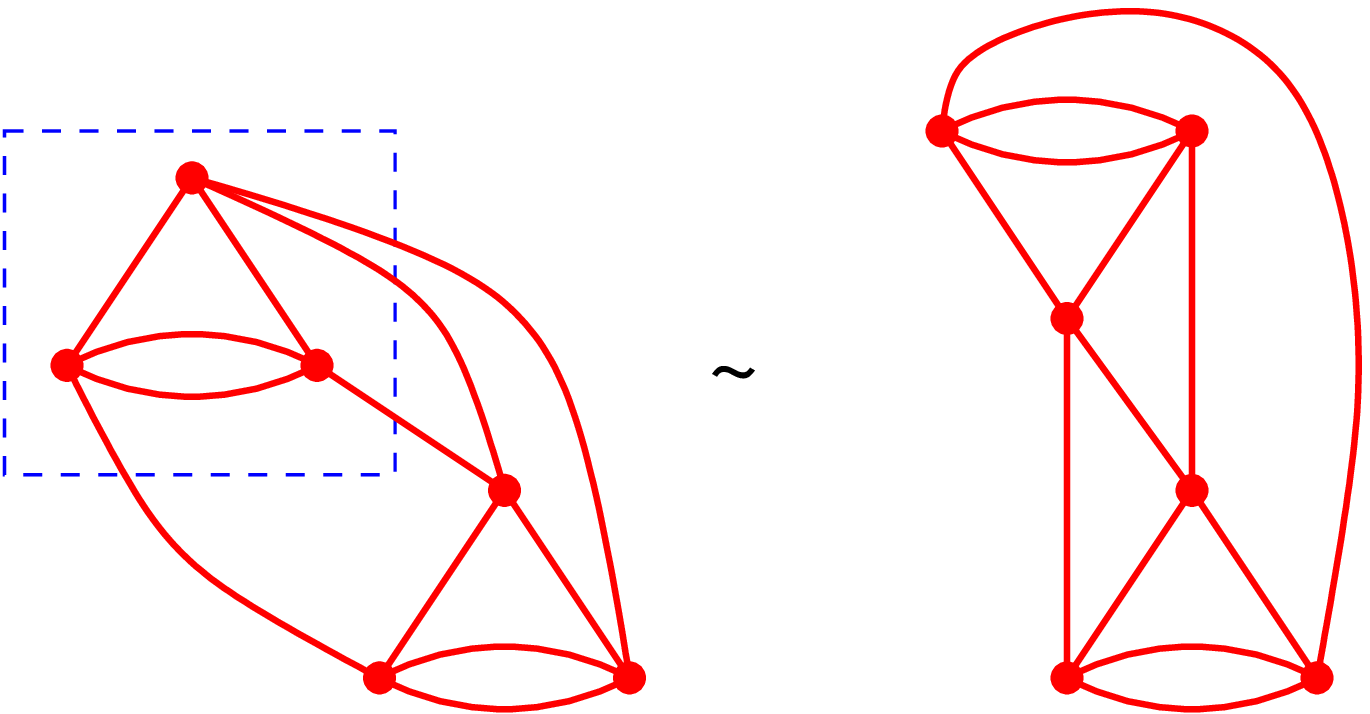}
\end{minipage}

The asymptotic behaviour 
{$F_p\sim {\rm const} \left({27/4}\right)^p p^{-7/2}$} 
exhibits the same ``critical exponent'' $-7/2$ as in (\ref{Fasymp1}) but an increased radius of convergence, 
as expected. Likewise $\gamma_p\sim {\rm const} \left({27/4}\right)^p p^{-5/2}$.

In a second step we must take the quotient by the flype equivalence. 
Sundberg and Thistlethwaite \cite{ST-tangles} proved that the flype equivalence
      can be dealt with by a suitable combinatorial analysis. The net result
       of their rigourous analysis is that the connected 4-point function
      $\tilde \Gamma$ can be deduced from $\Gamma(g)$ by a suitable change
      of variable: this final computation has been rephrased in  
      \cite{artic17}  where
     it is shown that it can be elegantly presented as a coupling
      constant renormalisation $g\to g_0$.
 In other words, 
 start from {$N \tr\left( \oh t M^2 -{{g_0}\over 4} M^4\right)$}, 
fix {$t=t(g_0)$} as before. Then compute  {$\Gamma(g_0)$}
and determine $g_0(g)$ as the solution of 
\be
 g_0 = g\left(-1 + {2\over (1 - g)(1 + \Gamma(g_0))}\right)\ ,
 \label{g00}
 \ee 
then 
the desired generating function is  {$\tilde\Gamma(g)=\Gamma(g_0)$}.

To show this, we introduce  
{$H(g)$}, the generating function of 
 ``horizontally-two-particle-irreducible'' (H2PI)  2-tangle diagrams, {\it i.e.} of 
diagrams whose left part cannot be separated from the right by cutting two lines.
Its Feynman diagram expansion reads 

\[
H=\vcenterbox{\includegraphics[width=.6\textwidth]{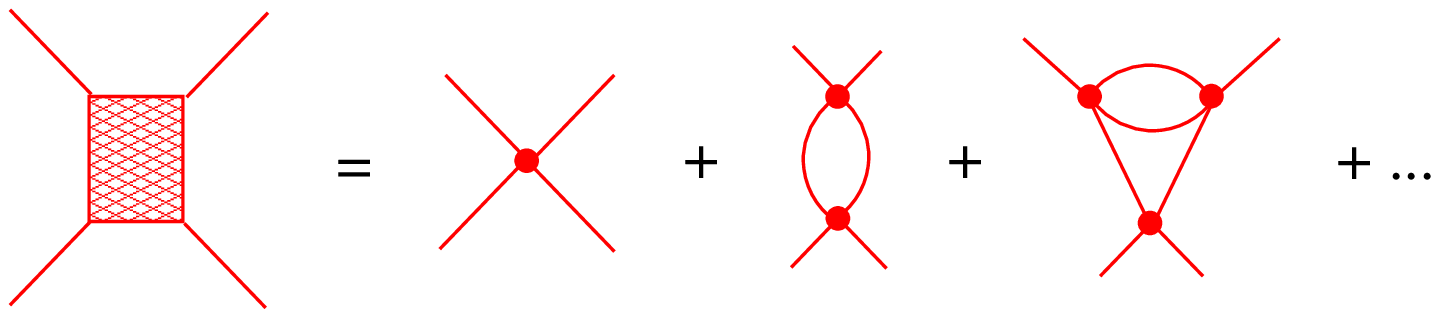}}
\]

\noindent Then the 4-point function $\Gamma$ is a geometric series of $H$  

\qquad  \includegraphics[width=.7\textwidth]{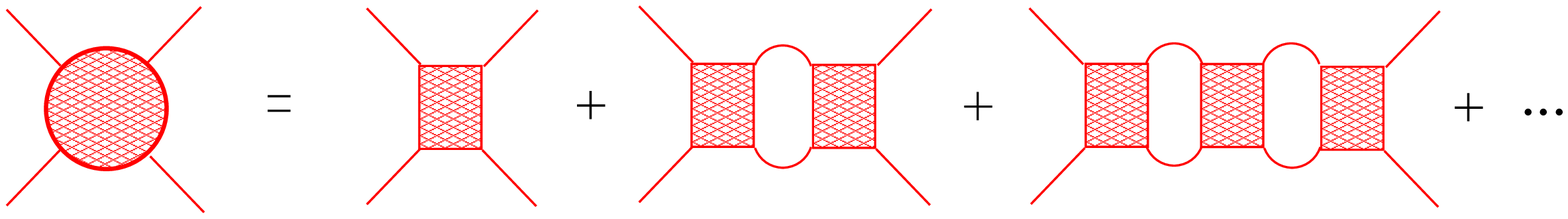}

\noindent summing up to {$\Gamma {=}H/(1-H)$}.

Now under the flype  equivalence  
 $\vcenterbox{ \includegraphics[width=.28\textwidth]{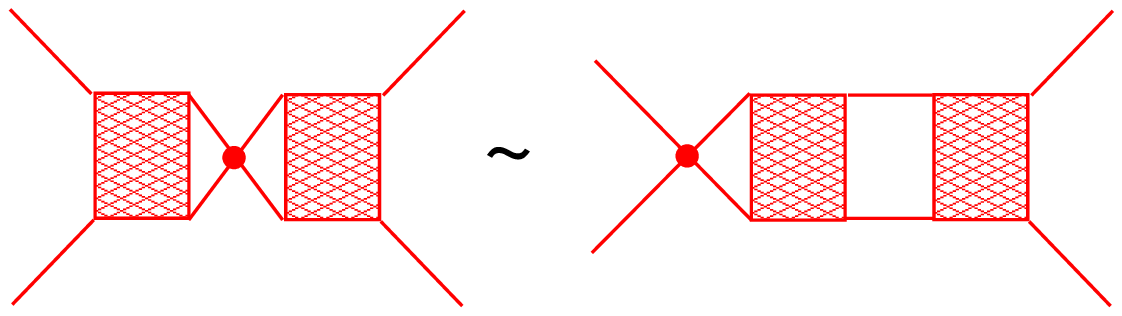}}$.
Thus, with $\tG$, resp. $\tH$ denoting generating functions of
{\it flype equivalence classes} of prime tangles, resp. of  H2PI tangles 
and if $\tH'$ is the non-trivial part of $\tH$, $\tH=g+\tH'$,
$\tilde\Gamma$ satisfies a simple recursive equation 
\begin{equation}
\tilde\Gamma=g+ g \tilde\Gamma +{\tH'\over 1-\tH'}\,,
\label{Gammatilde}
\end{equation}
both relations being depicted as
\[
\tilde H=
\vcenterbox{\includegraphics[width=0.35\textwidth]{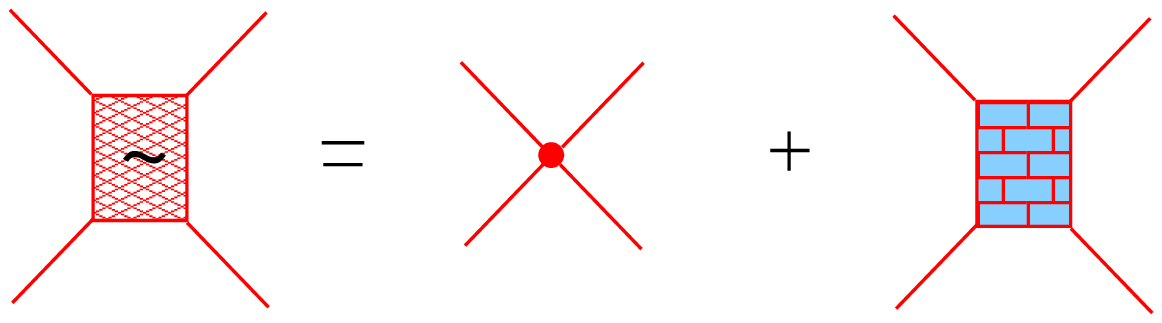}}
\]
\[
\tilde \Gamma=
\vcenterbox{\includegraphics[width=\textwidth]{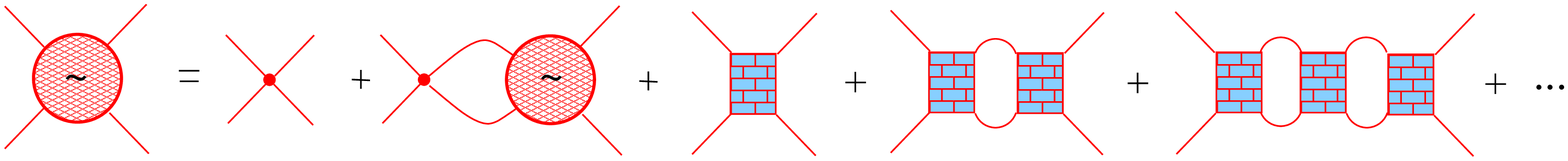}}
\]

Consider now the perturbative expansion of  $\Gamma(g_0)$ computed for a new value
$g_0$ of the coupling constant, depicted as an open circle
\[
\Gamma(g_0)=\vcenterbox{\includegraphics[width=.9\textwidth]{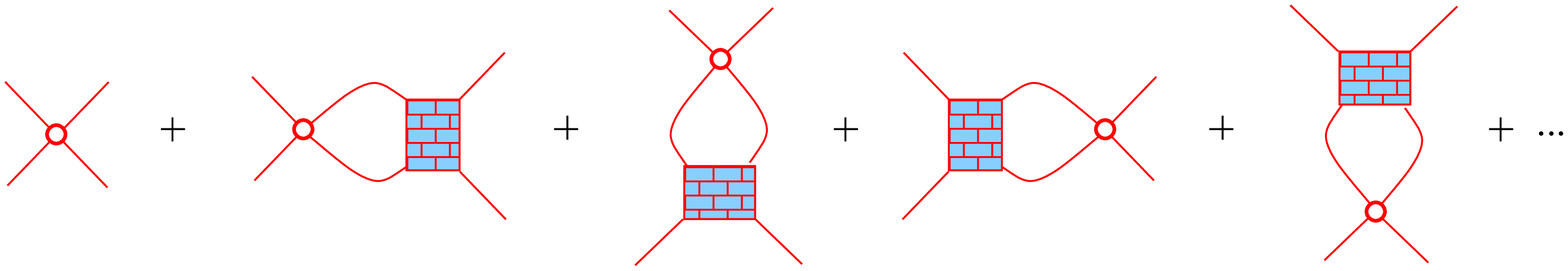}}
\]
If we want to identify it to $\tilde\Gamma(g)$, it is
suggested to determine $g_0=g_0(g)$ by demanding that $g_0 =g -2 g \tH'-\dots$ 
\[
g_0=\vcenterbox{\includegraphics[width=.7\textwidth]{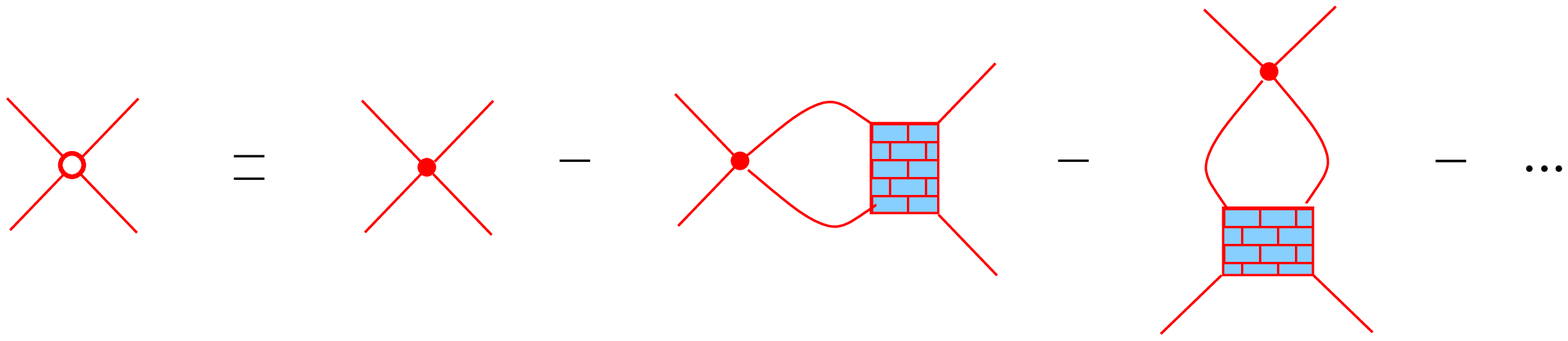}}
\]
so as to remove the first flype redundancies, 
and the remarkable point is that the ellipsis may be omitted and that no further term is required. 
Indeed  eliminating  {$\tH'$} 
between the two relations (\ref{Gammatilde}) and  $g_0 =g -2 g \tH'$ gives 
\be
g_0=g\Big(    -1 +\frac{2}{(1-g)(\tilde\Gamma+1)}\Big)
\label{g0}
\ee
which is equivalent to (\ref{g00}) and also 
to relations found in \cite{ST-tangles}.  
In the case of the matrix integral (\ref{inth}),
it is convenient to parametrize things in terms of 
$A=\frac{6}{4-a^2}$. One finds 
\begin{align}
\tilde\Gamma&= \frac{(A-2)(4-A)}{4}\\
 g_0&= \frac{4(A-2)}{A^3}\, ,
\end{align}
where $\tilde \Gamma$ is the wanted generating 
function of the number of flype-equivalence classes of prime alternating 2-tangles.
Eliminating $\tilde \Gamma$ and $g_0$ between 
the three latter equations 
results in a degree five equation for $A$
\be 
A^5 g-6 A^4 g+\frac{4 A^3 \left(g^2-2 g-1\right)}{g-1}-32 A^2+64 A-32=0
\ee
of which we have to find the solution which goes to 2 as $g\to 0$
$$ 
{A=2+2 g+6 g^2+20 g^3+78 g^4+334 g^5+1532 g^6+7372 g^7+36734 g^8+187902 g^9+\cdots 
}
$$
This then gives for $\tilde \Gamma$ the following expansion (given up to order 50 in \cite{ST-tangles})
\begin{equation}
\tG(g)=g+2g^2+4g^3+10g^4+29g^5+98g^6+372g^7+1538g^8+6755g^9
+\cdots
\label{Gammatildeexp}
\end{equation}
and the asymptotic behaviour of the $p$-th order of that expansion reads
\be \tilde\gamma_p\sim {\rm const}\ \left({101+\sqrt{21001}\over 40}
\right)^p p^{-5/2}\ee
with again the same exponent $-5/2$ but a still increased radius of convergence.

At this stage, we have merely reproduced the results of \cite{ST-tangles}.  
Our matrix integral approach has however two merits. It simplifies the combinatorics and 
recasts the quotient by flype equivalence in the (physically) appealing language 
of renormalization. 
For example using the results of 
 \cite{BIPZ}, one may easily compute the 
connected $2\ell$-function which counts the number of flype-equivalence classes of prime 
alternating $\ell$-tangles~\cite{artic17}
\begin{align*}
\Gamma_{2\ell} 
&= \frac{c_\ell}{\ell!} (A-2)^{\ell-1} (3\ell-2-(\ell-1)A)\\
c_{\ell+1}&=\frac{1}{3\ell+1} \sum_{\ell/2\le q\le \ell} (-4)^{q-\ell}\frac{(\ell+q)!}{(2q-\ell)!(\ell-q)!}
\end{align*}
and the numbers of 3- and 4-tangles up to 9 crossings are given by
\begin{align}\label{sixpt}
\Gamma_6 
&=3 g^2+14 g^3+51 g^4+186 g^5+708 g^6+2850 g^7+12099 g^8+53756 g^9+\cdots 
\\
\Gamma_8 
&=12 g^3+90 g^4+468 g^5+2196 g^6+10044 g^7+46170 g^8+215832 g^9+\cdots 
\end{align}
 Our approach also opens the route to generalizations in two directions:
\begin{itemize}
\item higher genus surfaces and ``virtual'' links.
\item  counting of ``coloured'' links, with a potential access to the still open problem of
disentangling knots from links.
\end{itemize}

This is what we explore in the next two sections.
\sect{Virtual knots}\label{virtualkn}

\subsection{Definition}
The large $N$ ``planar''  limit of the matrix integral (\ref{intc}) 
has been shown to be directly related
to the  counting of links and tangles. It is thus a natural question to wonder what the 
subleading terms in the $N^{-2}$ expansion of that integral, {\it i.e.} its
higher genus contributions, correspond to from the knot theoretic standpoint.
If one realizes that ordinary links and knots may always be deformed to live in a spherical
shell $S^2\times I$, where the interval $I$  is homeomorphic to $[0,1]$, one is ready to see
that higher genus analogues exist. 
In fact, these objects may be defined in two alternative ways. 

\def\Sigme{\mathbf{\Sigma}}
First, as just suggested, they are curves 
embedded  in a ``thickened'' Riemann surface 
$\Sigme:=\Sigma\times [0,1]$,  
modulo isotopy in $\Sigme$, {\em and} modulo orientation-preserving
homeomorphisms of $\Sigma$, 
{\em and} modulo addition or subtraction of empty handles. 

But one may also focus on the planar representations of these objects.
This leads to the concept of virtual knot diagrams \cite{Kauffman-virtual,KM-virtual}.
In addition to the ordinary under- and over-crossings, one must introduce a
new type of {{\it virtual}} crossing, which somehow represents the crossing of two
different strands 
that belong to different sides of the surface but are seen as 
crossing in the planar projection. Thus virtual knots diagrams are made of 
\includegraphics[width=.2\textwidth]{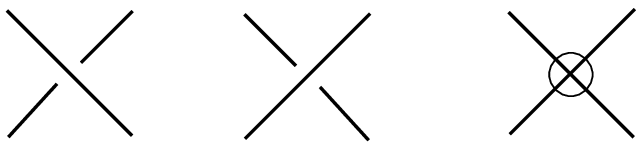}, and 
virtual links and knots are equivalence classes of such diagrams with respect to
the following generalized Reidemeister moves
\smallskip

\includegraphics[width=.9\textwidth]{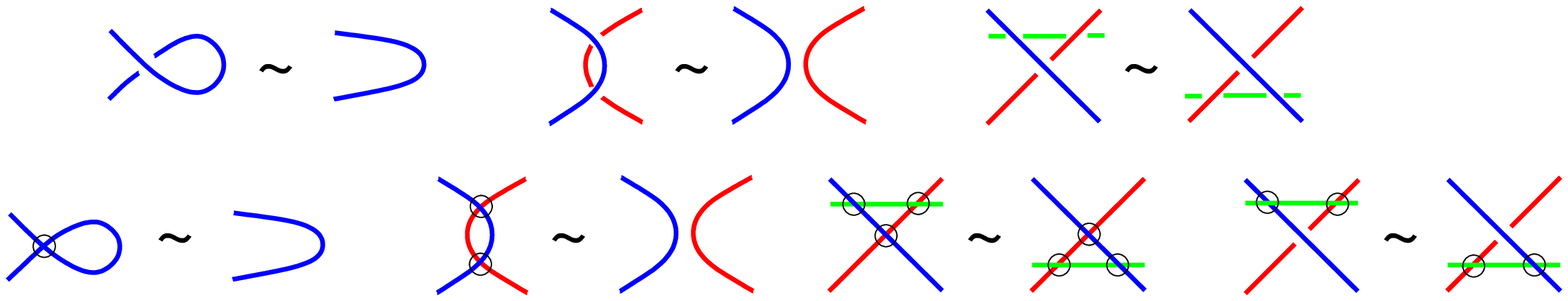}

\noindent That the two definitions are equivalent was proved in \cite{CKS-virtual,Kup-virtual}.
See \cite{Green-virtual} for a table of virtual knots.

Virtual alternating links and tangles are defined in the same way as in section 2: along 
each strand, one encounters alternatingly over- and under-crossings, paying 
no attention to possible virtual crossings.

\smallskip Here is a virtual link depicted in several alternative ways:
\smallskip

$\vcenterbox{\includegraphics[width=.8\textwidth]{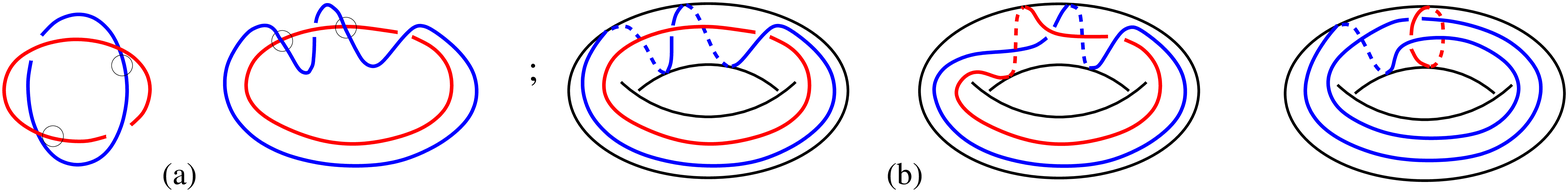}}$

\noindent 
in (a), using ordinary and virtual crossings; 
in (b), three equivalent 
representations on a Riemann surface.
 As illustrated by this example, in the thickened Riemann surface picture, 
the counting should be done irrespective of
the choice of homology basis or of the embedding of the link/knot. 
But this is precisely what higher genus 
Feynman diagrams of the matrix integral do for us!

{\it Remark:} there is a notion of genus for knots 
(minimal genus of a Seifert surface) which is unrelated to
the genus defined above (genus of the surface $\Sigma$). For the former
notion in the context of knot enumeration, see \cite{STV-virtual,SV-knots}.


\subsection{Higher genus contributions to integral (\ref{intc})}
We thus return to the integral (\ref{intc}) over {\it complex matrices}
{$$Z(g,t,N)=\int dM\, \E{N[-{t} \, \tr M \Md +{g\over 2} \tr (M \Md)^2]}
  $$}
and compute $F(g,t,N)=\frac{1}{N^2}\log Z(g,t,N)/Z(0,t,N)$ 
in an $N^{-2}$ expansion

 {$$   F(g,t,N)=\sum_{h=0}^\infty N^{2-2h} F^{(h)}(g,t)$$ }
$ F^{(h)}(g,t)$  receives contributions from Feynman diagrams of genus $h$.  
$F^{(0)}$ is (up to a factor 2) what was called $F$ in the previous section.
 $F^{(1)}$ was  computed in \cite{Morris}, 
$F^{(2)}$ and $F^{(3)}$ in \cite{Akemann-notes} and  \cite{Adamietz}. 
From $F$ one derives the expressions of $\Delta={1\over t} -{\partial F\over \partial t}$ and 
$\Gamma=2{\partial F\over \partial g}-2\Delta^2$.  
Moreover the first two terms in the  $g$ power series expansion of any $F^{(h)}(g)$ are 
easy to get \cite{artic25} and provide some additional information.

As before, we 
remove the non prime diagrams by imposing 
that $\Delta(g,t(g,N),N)$ $=1$, which determines {$t=t(g,N)$}  as a double $g$ and $1/N^2$ expansion. 
One then finds the generating function of prime 2-tangles of minimal genus $h$, 
$\Gamma^{(h)}(g)$, as the $N^{2-2h}$ term in the $1/N^2$ expansion of
$\Gamma(g,N)$:  
\vglue-5mm
\hglue-25mm
\begin{align*}
\scriptstyle{
\Gamma_{}^{(0)}(g) }
& \scriptstyle{
\,=\, g + 2\,g^2 + 6\,g^3 + 22\,g^4 + 91\,g^5 + 408\,g^6 + 
1938\,g^7 + 9614\,g^8 +   49335\,g^9 +260130\,g^{10}+\cdots
}\\ 
\scriptstyle{
\Gamma_{}^{(1)}(g) }
& \scriptstyle{ 
\,=\, g + 8\,g^2 + 59\,g^3 + 420\,g^4 + 2940\,g^5 + 20384\,g^6 
+ 140479\,g^7 +   964184\,g^8 + 6598481\,g^9
+45059872\,g^{10} +\cdots
}\cr 
\scriptstyle{
\Gamma_{}^{(2)}(g) }
& \scriptstyle{
\,=\, 17\,g^3 + 456\,g^4 + 7728\,g^5 + 104762\,g^6 + 1240518\,g^7 + 13406796\,g^8 + 
  135637190\,g^9 + 1305368592\,g^{10} +\cdots
}\\ \scriptstyle{
\Gamma_{}^{(3)}(g) }
& \scriptstyle{
\,=\, 1259\, g^5 + 62072\, g^6 + 1740158\, g^7 + 36316872\, g^8 + 
    627368680\, g^9 + 9484251920\,g^{10}  +\cdots
} \\  \scriptstyle{
\Gamma_{}^{(4)}(g) }
& \scriptstyle{
\,=\, 200589\ g^7 + 14910216\ g^8 + 600547192\ g^9
+ 17347802824\,g^{10}+\cdots
}\\ \scriptstyle{
\Gamma_{}^{(5)}(g) }
& \scriptstyle{
\,=\, 54766516\ g^9+ 5554165536\,g^{10}+\cdots
}
\end{align*}

\subsection{Table of genus 1, 2 and 3 virtual links with 4 crossings}

{\centering{\includegraphics[width=.95\textwidth]{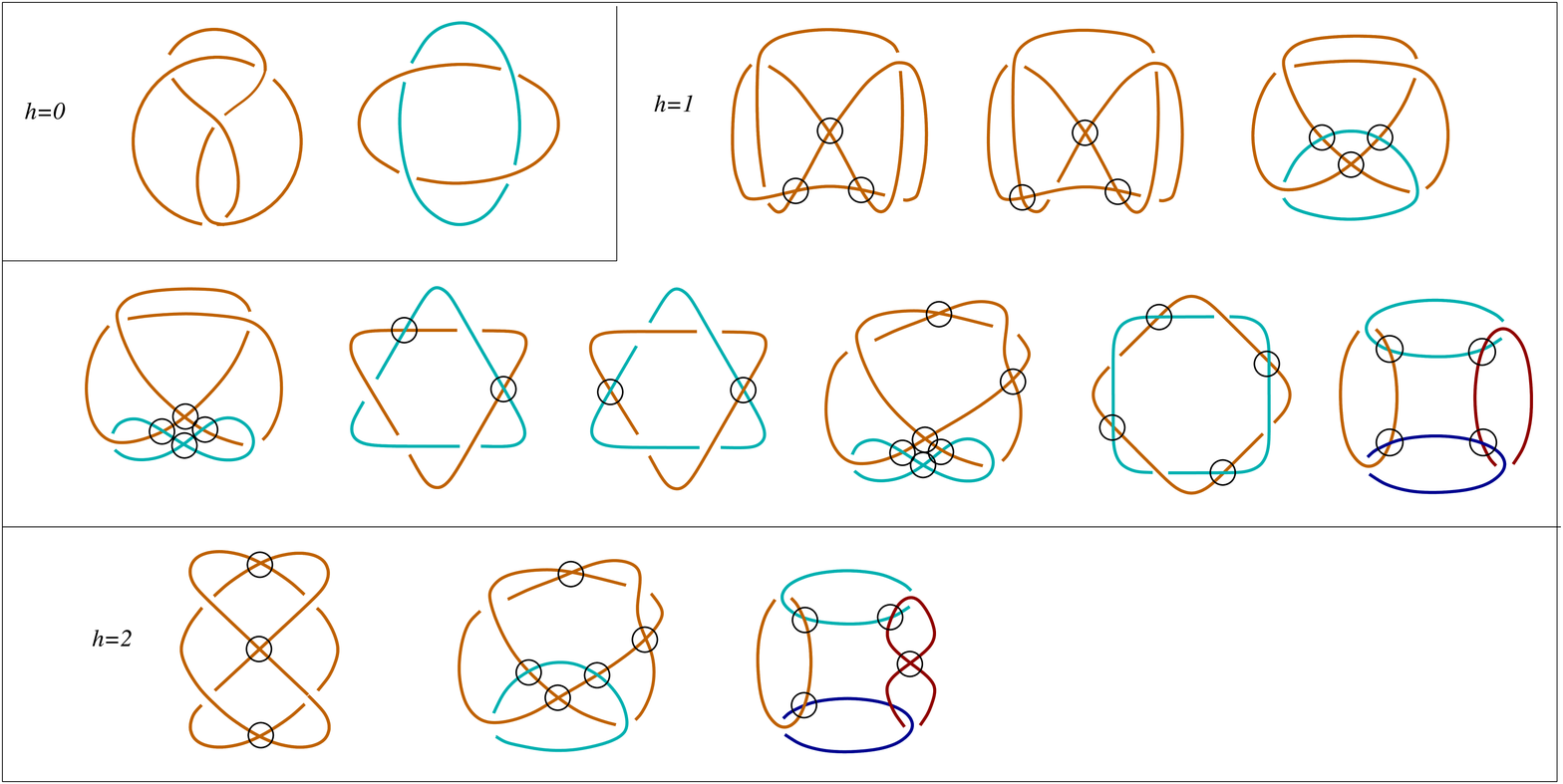}}}

Table of virtual knots and links with 4 crossings. Objects are not distinguished from their
mirror images, see \cite{artic25} for details.

\subsection{Removing the flype redundancies.}

The first occurences of flype equivalences occur in tangles with 3
crossings:
\includegraphics[width=.95\textwidth]{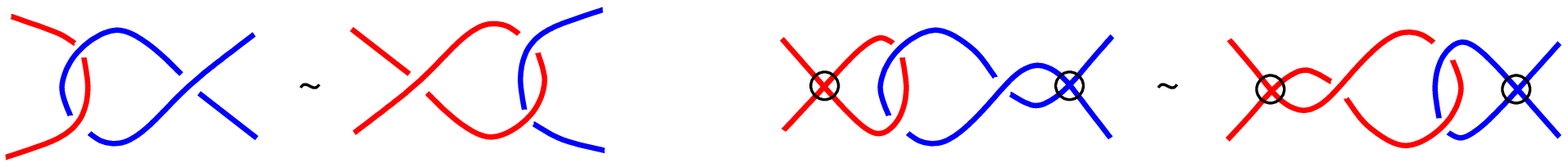}\\ 
It has been suggested \cite{artic25} that it is (necessary and) sufficient to
take the quotient by {\it planar} flypes, thus to 
perform the {\it same} renormalization {$g\to g_0(g)$}
as for genus 0. In other words, we have the 

{\bf Generalized flype conjecture:} \quad{\sl For a given (minimal) genus {$h$},  
{$\widetilde\Gamma^{(h)}(g)=\Gamma^{(h)}(g_0)$} is the 
generating function of flype-equivalence classes of virtual
alternating tangles.}

Then denoting by $\widetilde{\Gamma}^{(h)}(g)=\Gamma^{(h)}(g_0)$
 the generating function of the number of flype equivalence classes
of prime virtual alternating 2-tangles of minimal genus $h$,
$\widetilde{\Gamma}^{(0)}(g)$ is what was called $\widetilde{\Gamma}(g)$
in section 2, Eq. (\ref{Gammatilde}), while
\vskip-6mm
\begin{align*}  
 \scriptstyle{
 \widetilde{\Gamma}^{(1)}(g)}&=\  
 \scriptstyle{ g + 8\,g^2 + 57\,g^3 + 384\,g^4 + 2512\,g^5 
+ 16158\,g^6  
+ 102837\,g^7 + 649862\,g^8 + 
  4086137\,g^9+ 25597900\,g^{10} +\cdots
  } \\
 \scriptstyle{
 \widetilde{\Gamma}^{(2)}(g)}&=\ 
 \scriptstyle{17\,g^3 + 456\,g^4 + 7626\,g^5 + 100910\,g^6 + 1155636\,g^7 
+ 11987082\,g^8 +   115664638\,g^9+ 1056131412\,g^{10} +\cdots
} \\
 \scriptstyle{
 \widetilde{\Gamma}^{(3)}(g)}&=\  
 \scriptstyle{1259\,g^5 + 62072\,g^6 + 1727568\,g^7 + 35546828\,g^8 
+ 601504150\,g^9 +  8854470134\,g^{10}+\cdots
 } \\
 \scriptstyle{
 \widetilde{\Gamma}^{(4)}(g)}&=\  
 \scriptstyle{200589\ g^7 + 14910216\ g^8
 + 597738946\ g^9+ 17103622876\,g^{10}
 + \cdots
 }\\
 \scriptstyle{
 \widetilde{\Gamma}^{(5)}(g)}&=\  
 \scriptstyle{54766516\ g^9+ 5554165536\,g^{10} +\cdots
 }
\end{align*} 

 The asymptotic behavior of the number of  inequivalent tangles of order $p$ is
{$$
\ \tilde \gamma_p^{(h)}\sim    \left({101+\sqrt{21001}\over 40}\right)^p
p^{{5\over 2}(h-1)}\ .$$}

In \cite{artic25},  this generalized flype conjecture was tested
up to 4 crossings for links and 5 crossings for tangles by computing  as many 
distinct invariants of virtual links as possible. We refer the reader to that reference for a detailed discussion. No counterexamples were found.

\sect{Coloured links}
\newcommand\Sym{\mathcal{S}_{2k}}


\subsection{The bare matrix model}
Let us first describe the ``bare'' model that describes
coloured link diagrams. Since we are only interested in the dominant
order as the size of the matrices $N$ goes to infinity, 
we can consider, as was argued in section \ref{sect131}, a model
of Hermitean matrices (as opposed to the complex matrices that
were necessary in section \ref{virtualkn} for virtual tangles).

Let us fix a positive integer $\tau$ -- the number of colours --
and define the following measure on the space of $\tau$ Hermitean matrices
$M_a$:
\begin{equation}\label{baremm}
\prod_{a=1}^\tau d M_a\, \exp\left(N\Tr\Big(-\frac{1}{2}\sum_{a=1}^\tau M_a^2 + \frac{g}{4}
\sum_{a,b=1}^\tau (M_a M_b)^2\Big)\right)
\end{equation}
This measure has an $O(\tau)$ symmetry where the matrices $M_a$ are
in the fundamental representation of $O(\tau)$.

\hskip-5mm\begin{minipage}[t]{0.78\textwidth}
\noindent
Expansion in perturbation series of the constant $g$ produces the
following Feynman diagrams: they are fat graphs (planar maps)
with vertices of valence 4, in which the colours cross each other
at each vertex, see the figure. 
The summation over $O(\tau)$ indices produces a factor of $\tau$ for
every colour loop.
\end{minipage} \hfill
\vspace{-.4cm}
\begin{minipage}[t]{0.18\textwidth}\vspace{0pt}
\centering \includegraphics[width=0.95\textwidth]{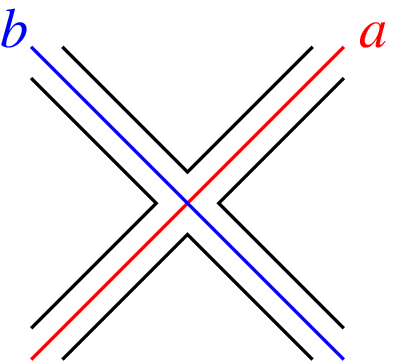}
\end{minipage}

\vspace{.4cm}
Thus, we have the following double expansion in $g$ and $\tau$:
\[
F=\lim_{N\to\infty}\frac{\log Z}{N^2}
=\!\!\!\!\sum_{\text{4-valent diagrams $D$}}
\frac{1}{|{\rm Aut} D|} 
\,
g^{\text{number of vertices}(D)} 
\,
\tau^{\text{number of loops}(D)}
\]
where it is understood that the number of loops is computed
by considering that colour loops cross each other at vertices.
In other words, the model of coloured links gives us more information than
the one-matrix model because it allows for a ``refined'' enumeration in which
one distinguishes the number of components of the underlying link.

Note that $F$ is at each order in $g$ a polynomial in $\tau$,
so that we can formally continue it to arbitrary non-integer values of $\tau$.
\subsubsection{Observables}\label{secobs}

Let $P_{2k}$ be the set of pairings of $2k$ points (sometimes called
``link patterns''), that is involutions of $\{1,\ldots,2k\}$
without fixed points. 

\begin{figure}
\centerline{%
\mbox{\includegraphics[scale=0.65]{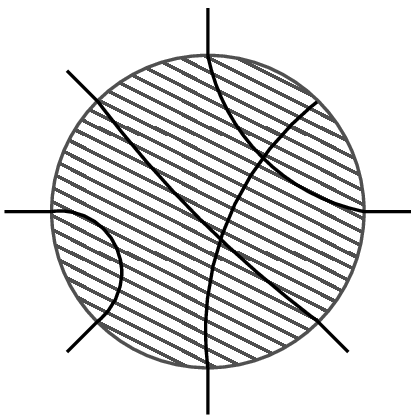}\break (a)}\qquad\qquad
\mbox{\includegraphics[scale=0.65]{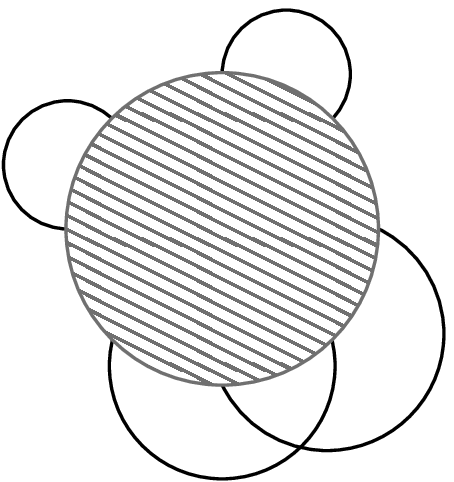}\break (b)}}%
\caption{The two types of correlation functions of the $O(\tau)$
matrix model.}\label{figconn}
\end{figure}

To each given link pattern $\pi$ of $2k$ points one can associate the quantity
$I_\pi$ ($I$ stands for ``internal connectivity'') as follows.
It is the generating series of the number of alternating $2k$-tangle diagrams
(or simply, of 4-valent fat graphs with $2k$ external legs)
with a weight of $\tau$ per closed loop and a weight of $g$ per vertex,
in such a way that the connectivity of the external legs,
which are numbered say clockwise from $1$ to $2k$,
is represented by $\pi$ (assuming as usual
that colours cross at each vertex).
See Fig.~\ref{figconn}(a).

From the point of view of the matrix model these observables
$I_\pi$ are not so natural. In principle one can define them as follows:
$$
I_\pi=\lim_{N\to\infty}\left\langle \tr M_{a_1}\ldots M_{a_{2k}} \right\rangle
\qquad a_i=a_j\ \Leftrightarrow\ j=i\ {\rm or}\ j=\pi(i)
$$
By $O(\tau)$ symmetry, the result is independent of the choice of the $a_i$
as long as they satisfy the condition above, i.e.\ that indices occur
exactly twice according to the link pattern $\pi$. However this
formula only makes sense if $\tau\ge k$.

A more natural quantity in the matrix model
is the ``external connectivity'' correlation function
$E_\pi$, which is defined in a very similar way:
$$
E_\pi=\lim_{N\to\infty}\sum_{a_1=1}^\tau\cdots\sum_{a_{2k}=1}^\tau
\prod_{i=1}^{2k} \delta_{a_i,a_{\pi(i)}}
\left\langle \tr M_{a_1}\ldots M_{a_{2k}} \right\rangle
$$
The only difference is that this time one sums over all $a_i$ (which might
produce additional coincidences of indices, and in fact always
will if $\tau<k$).

The graphical meaning of $E_\pi$ is that it is the
generating function of tangle diagrams with $2k$ external legs
and prescribed connectivity {\it outside}\/ the diagram, cf Fig.~\ref{figconn}(b).
Closing the external legs will produce closed loops which must be given a weight of $\tau$.
However, crossings outside the diagram should {\em not}\/ be given a weight of $g$.

Noting that 
all the diagrams that contribute to $E_\pi$ must have a certain internal
connectivity, we can write
\begin{equation}\label{exttoint}
E_\pi=\sum_{\pi'} G_{\pi,\pi'} I_{\pi'}
\end{equation}
The coefficients $G_{\pi,\pi'}$ are nothing but the natural 
scalar product on link patterns of same size $2k$, defined as
follows:
\[
G_{\pi,\pi'}=\tau^{\frac{1}{2}\text{number of cycles of $\pi\circ\pi'$}}
\qquad \pi,\pi'\in P_{2k}
\]
Graphically, it corresponds to gluing together the two pairings and giving a weight of $\tau$ to each closed loop that has been produced.

As a consequence of the formulae presented below,
for positive integer $\tau$ and $k>\tau$, the matrix $G$ has zero determinant
and formula \eqref{exttoint} cannot be inverted in the sense
that the $E_\pi$ are actually linearly dependent. For example at $\tau=1$
there is really only one observable per $k$ 
(with one colour one cannot distinguish
connectivities). It is however convenient to introduce the pseudo-inverse
$W$ of $G$, that is the matrix that satisfies $WGW=W$ and $GWG=G$.
The definition of $G$ still makes sense for non-integer $\tau$, in which
case $G$ is invertible and $W=G^{-1}$. We now sketch the computation of $W$
following \cite{CM-WOn}
(where it is called the Weingarten matrix, in reference
to \cite{Weingarten}). See also \cite{JBZ-On} for a recursive way to compute $W$ for generic $\tau$.

$G$ is a $(2k-1)!!\times(2k-1)!!$ symmetric matrix, with the property
that it is invariant by the action of the symmetric group,
where the latter acts on involutions by conjugation
($\sigma\cdot\pi=\sigma\pi\sigma^{-1}$):
$G_{\sigma\cdot\pi,\sigma\cdot\pi'}=G_{\pi,\pi'}$;
or equivalently $\sigma G=G\sigma$ for all $\sigma\in\Sym$.
Furthermore, one easily finds that
$\mathbb{C}[P_{2k}]$ contains exactly once every irreducible
representation of $\Sym$ associated to a Young diagram
with {\em even lengths of rows}.
Thus, $G$ is a linear combination of projectors onto these irreducible 
subrepresentations, which are of the form
\[
P^\lambda_{\pi,\pi'}=\frac{\chi^\lambda(1)}{|\Sym|}\sum_{\sigma\in\Sym: \sigma\cdot\pi'=\pi}
\chi^\lambda(\sigma^{-1})
\]
where $\lambda$ is a Young diagram with $2k$ boxes ($\lambda=2\mu$
for the projector $P^\lambda$ to be non-zero) and
$\chi^\lambda$ is the associated character of
the symmetric group.

Finally one can write
$
G=\sum_{\mu}
c_\mu
P^{2\mu}
$ where $\mu$ is a Young diagram with $k$ boxes, and
the coefficients $c_\mu$ can be computed \cite{CM-WOn,artic51}:
\begin{equation}\label{coeffjm}
c_\mu=\prod_{(i,j)\in\mu}
(\tau+2j-i-1)
\end{equation}

Therefore, the pseudo-inverse $W$ of $G$ can be written as
\begin{equation}\label{pseudoinv}
W=\sum_{\mu:\, c_\mu\ne 0} c_\mu^{-1} P^{2\mu}
\end{equation}

\subsubsection{Loop equations}
Loop equations are simply recursion relations satisfied by the correlation
functions of our matrix model. They can in fact be derived 
graphically without any reference to the matrix model, in which case
the parameter $\tau$ can be taken to be arbitrary (not necessarily a positive
integer). We recall that 
we limit ourselves to the dominant order as $N\to\infty$.

\begin{figure}
\[
\vcenterbox{\includegraphics[scale=0.65]{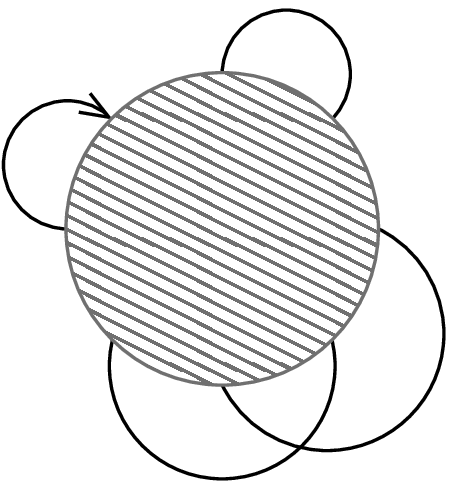}}
=
g \vcenterbox{\includegraphics[scale=0.65]{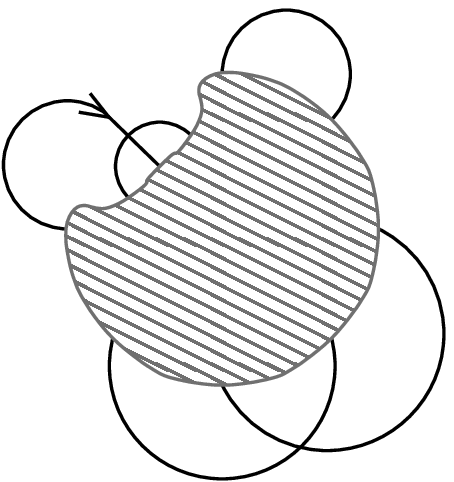}}
+
\sum_{i=1}^k\vcenterbox{\includegraphics[scale=0.65]{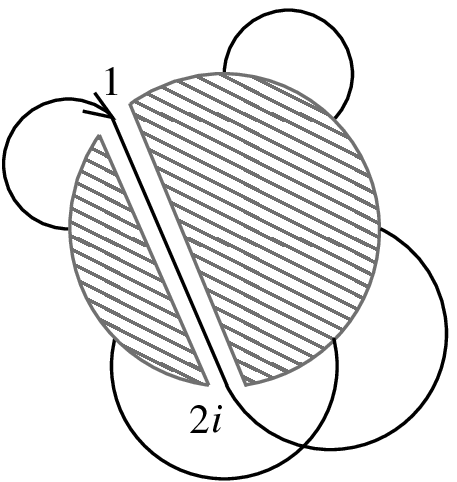}}
\]
\caption{Graphical decomposition for $E_\pi$.}\label{figloopeq}
\end{figure}

The recursion satisfied by $E_\pi$ is illustrated on Fig.~\ref{figloopeq}.
Start with one of the external legs (say leg numbered one),
and look at what happens to it once one moves inside the ``blob''.
There are two possibilities: (i) it reaches a crossing, in which case
one gets a factor of $g$ and a new correlation function $E_{\pi'}$
where $\pi'$ is obtained from $\pi$ by adding one arch around the leg number
one; or (ii) it goes out directly and connects to the external leg
$2i$, $i=1,\ldots,k$ (possibly creating
a loop and therefore a factor of $\tau$ if $\pi(1)=2i$). 
This second situation is more complex
because naively the two blobs created by cutting the initial blob into two
may still be connected by say $2\ell_i$ lines. Let us consider the two limiting
cases.
If $\ell_i=0$ we simply have two disconnected blobs and the contribution
is $E_{\pi_1} E_{\pi_2}$ where $\pi_1$ and $\pi_2$ are connectivities
of size $2(i-1)$ and $2(k-i)$.
On the contrary if $\ell_i=k-1$, the two blobs are fully connected to each
other according to a certain permutation $\sigma\in \mathcal{S}_{2(k-1)}$ 
and it is clear that internal connectivity for one becomes external
connectivity for the other, so that the contribution is of the form
$\sum_{\pi_1\in P_{2(k-1)}} E_{\pi_1} I_{\sigma\cdot\pi_1}$.
The crucial remark is that one can rewrite this as
$\sum_{\pi_1,\pi_2\in P_{2(k-1)}} W_{\sigma\cdot\pi_1,\pi_2}E_{\pi_1} E_{\pi_2}$ even
if $G$ is non-invertible. Indeed the $E_\pi$, due to formula \eqref{exttoint},
live in the image of $G$ and therefore one can ignore the zero modes
of $G$ ($G$ being symmetric, its image and kernel are orthogonal).
In the general case in which
there are $2\ell_i$ connections between the two blobs with associated
permutation $\sigma\in\mathcal{S}_{2\ell_i}$, one has
to break these connections by using
the matrix $W$ for link patterns of size $2\ell_i$;
calling $\pi_1(\rho_1)$ the connectivity of the first blob
in which the $2\ell_i$ legs connecting it to the other blob
have been replaced with the link pattern $\sigma^{-1}\cdot\rho_1$ of size $2\ell_i$,
and similarly for $\pi_2(\rho_2)$ and link pattern $\rho_2$, we get
an expression of the form
\begin{equation}\label{loopeq}
E_{\pi} = g E_{\pi'}+\sum_{i=1}^k
\tau^{\delta_{\pi(1),2i}}
\sum_{\rho_1,\rho_2\in P_{2\ell_i}}
W_{\rho_1,\rho_2} E_{\pi_1(\rho_1)} E_{\pi_2(\rho_2)}
\end{equation}
(where $W_{\emptyset,\emptyset}=E_{\emptyset}=1$).
This equation allows to calculate the $E_\pi$ iteratively, in the sense
that to compute the l.h.s. at a given order, 
the $E_\pi$ appearing in the r.h.s.\ are either needed at a lower order in $g$, 
or at the same order in $g$
but have fewer external legs, than the $E_\pi$ in the l.h.s.

\subsection{Removal of redundancies and renormalized model}
As in section \ref{sect133}, we now discuss how to go from the counting of
(coloured) alternating link diagrams to the counting of actual 
(coloured) alternating links, that is up to topological equivalences.
We recall that the process involves two steps: removal of nugatory crossings
and consideration of prime tangles only, which amounts to 
a wave function renormalization (i.e.\ renormalization of the quadratic
term of the action); and inclusion of flypes, which amounts to a renormalization
of the quartic term of the action. However, a crucial difference
with the model discussed in section \ref{sect133} is that in the $O(\tau)$ model
of coloured links one can introduce not just one, but two $O(\tau)$-invariant
quartic terms: besides the already present term of the form
$\Tr \sum_{a,b} (M_a M_b)^2$, one can also have another term of the form
$\Tr \sum_{a,b} M_a^2 M_b^2$, and one expects that this term will be
generated by the renormalization \cite{artic09}. We now summarize the equations
that we find. We start from the measure
\begin{equation}
\prod_{a=1}^\tau d M_a
\, \exp\left(N\Tr\Big(-{t\over 2} \sum_{a=1}^\tau M_a^2
+{g_1\over 4} \sum_{a,b=1}^\tau (M_a M_b)^2
+{g_2\over 2} \sum_{a,b=1}^\tau M_a^2 M_b^2
\Big)\right)
\end{equation}
The Feynman rules of this model now allow loops of different colours to
``avoid'' each other, which one can imagine as tangencies. The loop
equations of this model generalize in an obvious those of section xxx and will not be written here.

Next we define the following 
correlation functions $\Delta$ and $\Gamma_{0,\pm}$:
\begin{align*}
\Delta&=E_{(12)}/\tau\\
\Gamma_0&=(E_{(12)(34)}-\tau(\tau+1)\Delta^2)/n \\
\Gamma_\pm &= I_{(12)(34)}-\Delta^2\pm I_{(13)(24)}
\end{align*}
The $I_\pi$ are not directly defined in the matrix model, but the $E_\pi$ are
-- in fact the 4-point functions $E_\pi$ are obtained by differentiating
the free energy with respect to $g_1$ and $g_2$.
But from the formulae of section \ref{secobs} one can check that the Weingarten function
for 4-point functions is invertible for $\tau\ne 1,-2$ (special cases which
only require $\Gamma_0$, as discussed in detail in \cite{artic17}, 
and which we exclude
from now on). Thus one can deduce the 4-point $I_\pi$ from the $E_\pi$.
Also define the auxiliary objects (generating series
of horizontally two-particle irreducible diagrams)
\begin{align*}
H_0&=1-\frac{1}{(1-g)(1+\Gamma_0)}\\
H_\pm&=1-\frac{1}{(1\mp g)(1+\Gamma_\pm)}
\end{align*}

Then the equations to impose on the bare parameters $g_1$, $g_2$ and $t$
as functions of the renormalized coupling constant $g$ are
\begin{align}
\Delta&=1\\
g_1&=g(1-H_+-H_-)\\
g_2&=-g(H_0/\tau+(1/2-1/\tau)H_+-H_-/2)
\end{align}

Up to order 8, we find
\begin{align*}
\scriptstyle
g_1&\scriptstyle
\,=\,g-(2g^4+(2+2\tau)g^5+(14+2\tau)g^6+(26+16\tau+2\tau^2)g^7+(134+56\tau+2\tau^2)g^8)+\cdots\\
\scriptstyle
g_2&\scriptstyle
\,=\,-(g^3+g^4+3g^5+(5+2\tau)g^6+(27+5\tau)g^7+(89+32\tau+\tau^2)g^8)+\cdots\\
\scriptstyle
t&\scriptstyle
\,=\,1+2g+\tau g^2-2\tau g^3-6g^4-(8+10\tau)g^5-(38+16\tau+3\tau^2)g^6
\\&\hskip4.5cm\scriptstyle 
-(104+86\tau+14\tau^2)g^7-(410+338\tau+56\tau^2+2\tau^3)g^8+\cdots
\end{align*}

Composing these series with the correlation functions allows to produce
generating series for the number of coloured (prime) alternating tangles with
arbitrary connectivity. For example, we find for 4- and 6-tangles
(we only mention one pairing per class
of rotationally equivalent pairings):
\begin{align*}
\scriptstyle
I^c_{(12)(34)}&\scriptstyle
\,=\,g^2+g^3+(3+\tau)g^4+(9+\tau)g^5+(21+11\tau+\tau^2)g^6+(101+32\tau+\tau^2)g^7+\cdots\\
\scriptstyle
I^c_{(13)(24)}&\scriptstyle
\,=\,g+2g^3+2g^4+(6+3\tau)g^5+(30+2\tau)g^6+(62+40\tau+2\tau^2)g^7+\cdots\\
\scriptstyle
I^c_{(14)(25)(36)}&\scriptstyle
\,=\,2g^3+18g^5+18g^6+(156+24\tau)g^7+\cdots\\
\scriptstyle
I^c_{(14)(26)(35)}&\scriptstyle
\,=\,g^2+7g^4+6g^5+(53+8\tau)g^6+(154+6\tau)g^7+\cdots\\
\scriptstyle
I^c_{(12)(35)(46)}&\scriptstyle
\,=\,2g^3+2g^4+(16+2\tau)g^5+(42+2\tau)g^6+(171+44\tau+2\tau^2)g^7+\cdots\\
\scriptstyle
I^c_{(14)(23)(56)}&\scriptstyle
\,=\,4g^4+8g^5+(42+7\tau)g^6+(156+14\tau)g^7+\cdots\\
\scriptstyle
I^c_{(12)(34)(56)}&\scriptstyle
\,=\,3g^4+9g^5+(41+7\tau)g^6+(168+21\tau)g^7+\cdots
\end{align*}
The superscript $c$ means we are considering 
the {\em connected}\/ generating series (corresponding
to tangles which cannot be broken into several disentangled pieces)
e.g.\ $I^c_{(12)(34)}=I_{(12)(34)}-1$, $I^c_{(13)(24)}=I_{(13)(24)}$, etc.

In particular, 
note that $2I^c_{(12)(34)}+I^c_{(13)(24)}$ at $\tau=1$ reproduces Eq.~\eqref{Gammatildeexp},
and similarly $I^c_{(14)(25)(35)}+3I^c_{(14)(26)(35)}+6I^c_{(12)(35)(46)}+3I^c_{(14)(23)(56)}+2I^c_{(12)(34)(56)}$ at $\tau=1$ reproduces Eq.~\eqref{sixpt}.

\subsection{Case of $\tau=2$ or the counting of oriented tangles and links}
In the case of the (renormalized) matrix model with $\tau=2$ matrices, 
a subset of correlation functions can be computed
exactly, including the two- and four-point functions which are necessary
for our enumeration problem. Instead of giving two colours to each loop,
one can equivalently give them two orientations: not only does this give
a nice interpretation of the enumeration problem as the counting
of {\em oriented}\/ tangles, but it is also the first
step towards the exact solution of the problem. Indeed, as shown in
\cite{artic11},
this reduces it to the solution of the {\em six-vertex model}\/ on dynamical
random lattices, which was studied in \cite{artic10,Kostov-6v}.

The explicit generating series are given in terms of elliptic Theta functions
and will not be given here; even their asymptotic (large order) behavior
is somewhat non-trivial to extract, and we quote here the result of
\cite{artic11}:
if $\gamma_p$ is the $p^{\mathrm{th}}$ term of one of the four-point
correlation functions,
\[
\gamma_p{\buildrel p\to\infty\over\sim}
\mathrm{cst}\ g_c^{-p} p^{-2} (\log p)^{-2}
\]
where $g_c$ is
the closest singularity to the origin of these generating series;
$1/g_c \approx 6.28329764$. Though the latter number is non-universal,
the subleading corrections are; they correspond to a $c=1$ conformal field
theory of a free boson coupled to quantum gravity.

\subsection{Case of $\tau=0$ or the counting of knots}
A case of particular interest is the limit $\tau\to 0$ of the matrix
model \eqref{baremm}. This can be considered as a ``replica limit'' where one sends
the number of replicas to zero. Alternatively, the $\tau\to 0$ matrix model
can be written explicitly using a supersymmetric
combination of usual (commuting)
and of Grassmannian (anticommuting) variables, see \cite{artic17}.

The observables are defined as follows:
\[
\hat E_\pi = \lim_{\tau\to 0} (\frac{1}{\tau} E_\pi)
\]
that is they correspond to tangles which, once closed from the outside,
form exactly one loop (i.e.\ form knots as opposed to links).

The loop equations of the bare model become
\begin{equation}\label{loopeqzero}
\hat E_{\pi} = g \hat E_{\pi'}+\sum_{\substack{i=1,\ldots,k\\ 2i\ne \pi(1)}}
\ \sum_{\rho_1,\rho_2\in P_{2\ell_i}}
\hat W_{\rho_1,\rho_2} \hat E_{\pi_1(\rho_1)} \hat E_{\pi_2(\rho_2)}
\end{equation}
where $\hat W$ is the pseudo-inverse of
$\hat G=\lim_{\tau\to0}(\tau^{-1} G)$.
Note that according to 
(\ref{coeffjm},\ref{pseudoinv}), the factor $\tau^{-1}$ cancels the trivial
zeroes of $G$ at $\tau=0$. These zeroes are simple
for diagrams with $\lambda_3\le 1$; 
the remaining diagrams (in size $n\ge 6$) have higher zeroes, making 
$\hat G$ non-invertible.

Though Eqs.~\eqref{loopeqzero} cannot be solved analytically, it is
worth mentioning that they are easily amenable to an iterative solution
by computer; in fact the resulting algorithm is notably better
than the transfer matrix approach of \cite{artic15,artic16}, 
and one finds for example for
the two point function $\Delta=\hat E_{(12)}$ the following power series in $g$:
(using a PC with 8 Gb of memory and 24h of CPU)
\begin{gather*}
\scriptstyle
1, 2, 8, 42, 260, 1796, 13396, 105706, 870772, 7420836, 65004584,
582521748, 5320936416, \\
\scriptstyle
49402687392, 465189744448, 4434492302426, 
42731740126228, 415736458808868,\\ 
\scriptstyle
4079436831493480, 40338413922226212, 
401652846850965808, 4024556509468827432,\\ 
\scriptstyle
40558226664529024000,
410887438338905738908, 4182776248940752113344,\\
\scriptstyle
42770152711524569532616, 439143340987014152920384, 
4526179842103708969039296\ldots
\end{gather*}
The objects being counted by this formula are also known as
self-intersecting plane curves or long curves, see e.g.\ \cite{GZD-planecurves}.

One can similarly take the limit $\tau\to 0$ in the renormalized model.
However little is known beyond the general facts mentioned above for
arbitrary $\tau$.

\subsection{Asymptotics}
The most interesting unsolved question about the $O(\tau)$ matrix model
of coloured links and tangles concerns the large order behavior of the
generating series in the coupling constant $g$, i.e.\ the asymptotic
number of coloured alternating tangles as the number of crossing is sent to infinity.
If one considers, in the spirit of chapter \rem{?}, 
that the model represents a statistical model on random lattices, then
it is expected that the model is critical for $|\tau|<2$ and non-critical
for $|\tau|>2$. This should affect the universal subleading power-law
corrections to the asymptotic behavior.

In \cite{artic24}, the following conjecture was made. For $|\tau|<2$,
the model corresponds to a theory with central charge $c=\tau-1$
(corresponding to the analytic continuation of a model of $\tau-1$ free
bosons). This implies the following behavior
for the series $\sum_p\gamma_p g^p$ counting coloured prime alternating tangles
(with, say, four external legs):
\[
\gamma_p(\tau){\buildrel p\to\infty\over\sim}
\mathrm{cst}\ g_c(\tau)^{-p} p^{\gamma(\tau)-2}
\qquad \gamma(\tau)=\frac{\tau-2-\sqrt{(2-\tau)(26-\tau)}}{12}
  \qquad |\tau|<2
\]
This was tested numerically in \cite{artic24}, but the results are not
entirely conclusive (see also \cite{artic16,artic29}).

In particular, as a corollary of the conjecture above, one would have
the following asymptotic behavior for the number of prime alternating
knots:
\[
f_p{\buildrel p\to\infty\over\sim}
\mathrm{cst}\ g_c(0)^{-p} p^{-\frac{19+\sqrt{13}}{6}}
\]
It is most likely that one can remove the ``prime'' property without changing
the form of the asymptotic behavior (only the non-universal coefficient of the 
exponential growth would be modified); one can speculate that removing the ``alternating''
property will not change it either.

Note the similarity between our problem of counting knots with that of
counting meanders \cite{DFGG-meanders}. There too, the problem can be rewritten as a matrix model
and the asymptotic behavior is dictated by 2D quantum gravity, leading to a non-rational
critical exponent. A key difference is that in the case of meanders, corrections to
the leading behaviour are expected to be power-law, making numerical checks reasonably easy.
In contrast, if the conjecture above for knots is correct, the corrections are expected
to be logarithmic (the theory being asymptotically free in the infra-red), which would make
numerical checks extremely hard.

{\sc Acknowledgements}:
This work was supported in part by 
EU Marie Curie Research Training Network
``ENRAGE'' MRTN-CT-2004-005616, ESF program ``MISGAM''
and ANR program ``GRANMA'' BLAN08-1-13695.

\newcommand\MR[1]{\relax\ifhmode\unskip\spacefactor3000 \space\fi
  \MRhref{#1}{{\sc mr}}}
\gdef\MRSIMPLIFY#1 #2MREND{#1}%
\newcommand{\MRhref}[2]{%
 \href{http://www.ams.org/mathscinet-getitem?mr=\MRSIMPLIFY#1 MREND}{#2}}
\DeclareUrlCommand\path{\urlstyle{tt}}
\bibliography{../biblio}
\bibliographystyle{../amsplainhyper}

\end{document}